\documentstyle[newarcrc,fleqn]{article}

\input{psfig.sty}

\title{The transition phase in Nova Muscae 1998, the dust formation in 
classical novae and a possible link to intermediate polar systems.}

\author{Retter A.\address{Department of Astronomy, Keele University, \\
        Staffs., ST5 5BG, UK; ar@astro.keele.ac.uk}%
        Liller W. \address{Vina Del Mar, Chile; wliller@compuserve.com}
        Garradd G.\address{Loomberah, N.S.W.;loomberah@ozemail.com.au}}

\begin{document}
\maketitle

\begin{abstract}
We present results of continuous CCD photometry of Nova Muscae 1998 
during 17 nights obtained mainly during the transition phase in the light 
curve of the young nova. We discovered a periodical oscillation with the 
period 0.16930 day. This result is important for understanding this 
peculiar stage in the light curve of certain classical novae. We also note 
a possible connection between the transition / dust phase in nova light 
curves and intermediate polar systems.
\end{abstract}

\section{Introduction}

Nova Muscae 1998 was discovered on 1998 December 29 (Liller 1998)
at $m_{V}$=8.5. The light curve of the nova during the first 100 days 
followed outburst, which was compiled from various sources is shown in 
Fig. 1. We estimate that the time to decline by two magnitudes is: 
$t_{2(V)}=3.5\pm0.3$ day. The object is therefore classified as a very 
fast nova.

Fig. 1 also displays oscillations with time scales of several days, and 
full amplitude of about one magnitude in the light curve of the nova 
during days 20-60 after maximum light. This phase of the nova is known 
as the transition phase. It is unclear why certain novae experience this
behaviour in their light curves (Leibowitz 1993).

We observed Nova Mus 1998 during 17 nights in 1999. The observations
were carried out using the 20-cm telescope in Vina Del Mar, Chile, and 
the 45-cm telescope in Loomberah, Australia. CCDs and broad band filters 
were used. Most of the observations were obtained during the transition 
phase of the nova. The highest peak in the power spectrum (not shown) 
corresponds to the period 0.16930 day +/-0.00015. 

\section{Discussion}

In the power spectrum there are a few more peaks with possible 
inter-connections, however they are below the significance level. The nova 
might thus be classified as an intermediate polar candidate.

In Table 1. we summarize the properties of all novae, which are 
intermediate polar systems or intermediate polar candidates. Although the
information is scarce, the data suggest that there might be a connection 
between the two groups. This link may be related with the evidence for 
the early presence of the accretion disc in young novae (Retter 1999). 


\begin{figure}




\psfig{figure=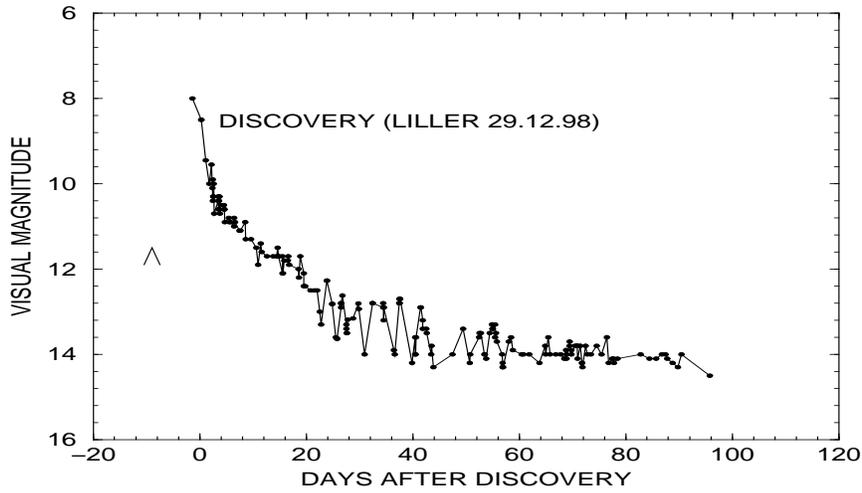,width=4.5in,height=2.6in}



\caption{Amateurs light curve of Nova Muscae 1998 over the first 
100 days after discovery.}


\end{figure}









%
%



\begin{table}[hbt]

\caption[]{Intermediate Polar Novae and Dust / Transition Phase}

\begin{tabular}{@{}ccccc@{}}

Nova     & Year &Intermediate Polar  & Dust Phase   & Transition Phase\\
\hline
GK Per   & 1901 &    +          &              &      +          \\
V603 Aql & 1918 &    +?         &              &      +          \\
DQ Her   & 1934 &    +          &      +       &                 \\
HZ Pup   & 1960 &    +?         &              &      ??         \\
V705 Cas & 1993 &    ??         &      +       &                 \\
V1425 Aql& 1995 &    +?         &prediction (from IR)&  ??        \\
Mus      & 1998 &    ??         &              &      +           \\
\hline

\end{tabular}

\end{table}

\end{document}